\documentclass[aps,floatfix,preprint,showpacs,preprintnumbers,nofootinbib,superscriptaddress,natbib]{revtex4}

\usepackage{graphicx,float}
\usepackage{epsfig}		
\usepackage{dcolumn}
\usepackage{bm}
\usepackage{amsmath,upgreek}
\usepackage{amssymb}
\usepackage{braket}
\usepackage{natbib}
\usepackage{letltxmacro} 

\usepackage[all]{xy}
\usepackage{pdfpages}
\usepackage{color}
\usepackage[colorlinks=true,
 linkcolor=red,
 urlcolor=purple,
 citecolor=blue,hyperindex]{hyperref}

\usepackage{braket}

\def\bkR{{\rm I\kern-.17em R}}

\def\0{\mbox{\tiny $0$}}
\def\1{\mbox{\tiny $1$}}
\def\2{\mbox{\tiny $2$}}
\def\3{\mbox{\tiny $3$}}
\def\4{\mbox{\tiny $4$}}
\def\5{\mbox{\tiny $5$}}
\def\6{\mbox{\tiny $6$}}
\def\7{\mbox{\tiny $7$}}
\def\8{\mbox{\tiny $8$}}
\def\9{\mbox{\tiny $9$}}

\def\f14{\mbox{\tiny $\frac{1}{4}$}}

\begin{document}

\title{Relativistic dispersion relation and putative metric structure in noncommutative phase-space}

\author{P. Leal}
\email{pedro.leal@fc.up.pt}
\affiliation{Departamento de F\'isica e Astronomia and Centro de F\'isica do Porto, Faculdade de Ci\^{e}ncias da
Universidade do Porto, Rua do Campo Alegre 687, 4169-007, Porto, Portugal.}
\author{O. Bertolami}
\email{orfeu.bertolami@fc.up.pt}
\affiliation{Departamento de F\'isica e Astronomia and Centro de F\'isica do Porto, Faculdade de Ci\^{e}ncias da
Universidade do Porto, Rua do Campo Alegre 687, 4169-007, Porto, Portugal.}

\begin{abstract}
The deformation of the relativistic dispersion relation caused by noncommutative (NC) Quantum Mechanics (QM) is studied using the extended phase-space formalism. The introduction of the additional commutation relations induces Lorentz invariance violation. It is shown that this deformation does not affect the propagation speed of free massless particles. From the deformation of the dispersion relation for massless particles, gamma ray burst data is used to establish an upper bound on the noncommutative  parameter, $\eta$, namely $\sqrt{\eta} \lesssim 10^{-12} \,\mathrm{eV/c}$. Additionally, a putative metric structure for the noncommutative phase-space is discussed.
\end{abstract}

\date{\today}
\maketitle

\maketitle

\section{Introduction}

Lorentz symmetry is one of the foundational principles of modern physics and, in particular, of Special and General Relativity. However, in the attempt to reach a quantum theory of gravity, several proposed theories admit some violation of this symmetry, by means of a deformation of the relativistic dispersion relation (see Refs.~\cite{Many_authors} for reviews). This takes place, for instance, in solutions of string theory \cite{Kostelecky_1989,Kostelecky_1991} and in noncommutative field theories \cite{Seiberg_1999,Caroll_2001,Bertolami_2003}. In a similar way, noncommutative (NC) quantum mechanics breaks this symmetry explicitly by introducing additional commutation relations, which provide an intrinsic momentum and length scale to the theory. As a consequence, this would lead to a deformation of the dispersion relation due to such scales. This may provide some useful insight into the deep consequences of NCQM. 

Phase-space NCQM (PSNCQM) is characterized by the introduction of momentum-momentum and position-position non-vanishing commutation relations and may be regarded as a deformation of the standard Heisenberg-Weyl (HW) algebra of these observables:
\begin{equation}
\left[\hat{q}_i,\hat{k}_j\right]=\mathrm{i}\hbar\delta_{ij}, \quad
\left[\hat{q}_i,\hat{q}_j\right]=\mathrm{i}\theta_{ij}, \quad \left[\hat{k}_i,\hat{k}_j\right]=\mathrm{i}\eta_{ij},
\end{equation}
where $\hat{q}$ and $\hat{k}$ are position and momentum operators, respectively, and $\theta_{ij}$ and $\eta_{ij}$ are antisymmetric matrices.

Alternatively, quantum mechanics may be viewed in the Wigner-Weyl (WW) phase space formalism, which is equivalent to the standard operator-state formalism \cite{Groenewold_1946,Moyal_1949,Wigner_1932}. As has been shown in Ref.~\cite{Bastos_2008}, a noncommutative extension of this formalism is possible and, besides being fully consistent, it is equivalent to the deformation of the HW algebra. This formalism embeds the quantum effects on the defined $\star$-product, which depends on the symplectic form of the phase-space. The introduction of the additional commutation relations in PSNCQM produces a deformation of this symplectic form, which captures all effects of NCQM through a modified Moyal product. This suggests that the effect of PSNCQM is to alter the geometric symplectic structure of the phase-space manifold and this will be used in the present work in order to compute the deformation in the relativistic dispersion relation. Our results suggest that there is a deformation of this relation that is suitable for phenomenological testing and, therefore, it might allow for constraining one of the noncommutative parameters of the theory.

Several of the effects of PSNCQM have been studied. Among these, implications for the uncertainty principles \cite{Bastos_2008_PRD,Bastos_2010_PRA,Bastos_2014,Bastos_2015}, for entanglement \cite{Bastos_2013_PRD} and decoherence \cite{Bernardini_2013_PRA,Santos_2015,Bernardini_2015} are the most conspicuous. The effect of a deformation of the relativistic dispersion relation based on the NC parameters has also been studied previously in Refs. \cite{Leal_2015,Queiroz_2011}. More recently, the implications of PSNCQM for the no-cloning theorem and for quantum teleportation were investigated \cite{Leal_2018}.

The outline of this paper is as follows. In Section II the extended phase-space formalism is introduced and motivated. The content of this Section is inspired on the formalism of Refs.~\cite{Struckmeier_2005,Struckmeier_2008}. The use of this formalism for the treatment of Lorentz transformations in phase-space (rather than in configuration space) is highlighted together with its use for obtaining relativistic Hamiltonians. In Section III the geometric implications of PSNCQM are introduced through the deformed symplectic form of the phase-space manifold. Together with the Darboux transformation, this is used to derive the deformed dispersion relation in the NC scenario. Additionally, it is shown that this deformation of the relativistic dispersion relation does not affect the speed of light for massless particles, in contrast with other deformations. Section IV is devoted to an astrophysical test of the deformed relation via the use of gamma ray burst (GRB) data and an upper bound on a NC parameter is obtained. Finally, Section V concerns the introduction of a metric structure on the phase-space via a compatible triplet that admits a symplecic form, an almost complex structure (ACS) and a metric tensor. In Section VI, our concluding remarks are drawn.

Throughout this work the following notation will be used. The two coordinate systems for phase-space will be denoted by $z=\{x,p\}$ and $\tilde{z}=\{q,k\}$. Coordinate indices for $z$ and $\tilde{z}$ will be used in the superscript form as well as for the coordinate indices of $x$ and $q$. On the other hand, indices for $p$ and $k$ will be denoted as subscripts.

\section{Extended phase-space formalism}

NCQM, likewise QM itself, is inherently non-relativistic, and thus it is not invariant under Lorentz transformations, but rather under Galilean transformations. However, it is logical to expect that its additional commutation relations imply some deformations of the relativistic dispersion relation \cite{Amelino_et_al_2015,Alexander_2010}. At first, one may think that since the HW algebra is deformed in a nonrelativistic context, it has little to do with relativistic physics. However, the introduction of the noncommutative parameters introduces in the theory a fundamental length and momentum scale. The existence of these scales is a fundamental aspect of the theory and can, in fact, be extended to the relativistic regime. In order to see this, we will not be directly concerned with PSNCQM itself, but rather with the deformation it introduces on the symplectic structure of the phase-space manifold. This discussion is based on the deformation caused in the standard symplectic structure of the WW formulation of quantum mechanics. Therefore,  we shall study the geometric properties of the phase-space as a symplectic manifold and assess the consequences of deforming this structure. Furthermore, since quantum mechanical phase-space is still non-relativistic (since it treats time in a special way), an approach is needed to symplectic manifolds that extends the usual notion of phase-space to incorporate time as a coordinate on this manifold rather than an external parameter. For this, we shall introduce the notion of an extended phase-space in the following section. Subsequently, the extended phase-space, which can be shown to have Lorentz invariance under certain conditions, may be deformed and the consequence of this deformation is the breaking of Lorentz symmetry and hence a correction to the dispersion relation arises. 

\subsection{Construction} \label{setup}
Consider the configuration space, a manifold, $M$, with coordinates $\{x^i\}$, $i=1,2,3$, and $T^*M$, the cotangent bundle of $M$, with coordinates $z=\{x^i,p_i\}$. Thus, a symplectic manifold of dimension $2n$ can be built, $(T^*M,\omega)$, where $\omega$ is a symplectic form defined on $T^*M$. It is always possible to define the symplectic form (at least locally) as the exterior derivative of some one form, i.e., 
\begin{equation}
\omega=\mathrm{d}\theta.
\end{equation}
Additionally, let us define an Hamiltonian, $H$, on $T^*M$, as $H: T^*M\rightarrow \mathbb{R}$. This leads to an Hamiltonian system, $(T^*M,\omega,H)$. If this Hamiltonian is an explicit function of time, i.e. $\partial_t H\neq 0$, then, it is useful to study the system in the manifold $T^*M\times \mathbb{R}$, with coordinates $\{p_i,x^i,t\}$, where the Poincaré-Cartan invariant form is defined as:
\begin{equation}\label{poincare_cartan_invariant}
\theta'=\theta-H(p_i,x^i,t)\,\mathrm{d}t. 
\end{equation}
This structure on $T^*M\times \mathbb{R}$ does not lead to a symplectic structure, since all symplectic manifolds must be even dimensional. One may thus think of expanding this structure with an aditional coordinate, say $p_0$, and define an extended phase-space. The $2n+1$-dimensional space $T^*M\times \mathbb{R}$ would be recovered under some on-shell condition for some function defined on the extended phase-space. This construction is indeed possible and a thorough explanation is provided in Refs. \cite{Struckmeier_2005,Struckmeier_2008}. An abridged version will be given in the following.

Consider a second manifold $M_1$ with coordinates $\{x_1^\mu\}=\{t,x^i\}$ and its tangent bundle $T^*M_1$ with coordinates $\{p^1_\mu,x_1^\mu\}=\{p_0,p_i,t,x^i\}$. This tangent bundle is known as the extended phase-space. An Hamiltonian, $H_1$, can now be defined on $T^*M_1$. However, since the objective is to describe a physical system on $T^*M\times \mathbb{R}$, some condition must be imposed on the extended Hamiltonian, $H_1$, and the appropriate coordinate $p^1_0$ must be defined. These requirements are met if one defines $p^1_0:=-\mathcal{H}$, where $\mathcal{H}$ is the value of the non-extended Hamiltonian, $H$. One should note the distinction between $H=H(p_i,t,x^i)$, which is a function on $T^*M\times \mathbb{R}$, and $\mathcal{H}$ which is the value this function assumes on some phase-space point. If one considers $s$ as a parameter on $T^*M_1$ (in the same way that $t$ is a paramenter on $T^*M$ for autonomous systems), then $\mathcal{H}=\mathcal{H}(s)=H(p_i(s),t(s),x^i(s))\in \mathbb{R}$. Since it only depends on the parameter $s$, then it is a suitable coordinate for $p^1_0$. As for the condition on $H_1$, one demands that $H_1=0$ generates a $(2n+1)$-dimensional hypersurface on $T^*M_1$ that matches the Hamiltonian system in $T^*M\times \mathbb{R}$. Thus, one can define a map $\Psi$: $T^*M\times \mathbb{R}\rightarrow T^*M_1$ such that:
\begin{equation}
\{t,x^i,p_i\}\rightarrow\{t,x^i,\mathcal{H},p_i\}:=\{x_1^\mu,p^1_\mu\}.
\end{equation}
The inverse map, $\Psi^{-1}:T^*M_1\rightarrow T^*M\times \mathbb{R}$, takes us back to the original system by replacing all instances of the $\mathcal{H}$ coordinate with the Hamiltonian $H$. The symplectic form on $T^*M_1$ is, in canonical coordinates, given by:
\begin{equation} \label{extended_sf}
\omega_e=\mathrm{d}p^1_\mu\wedge\mathrm{d}x_1^\mu=\mathrm{d}p_i\wedge\mathrm{d}x^i-\mathrm{d}\mathcal{H}\wedge\mathrm{d}t.
\end{equation}
If one applies $\Psi^{-1}$ to this form, one gets:
\begin{equation}
\omega_{e|_{\mathcal{H}=H}}=\mathrm{d}\theta',
\end{equation}
where $\theta'$ is the Poincaré-Cartan invariant defined in $T^*M\times\mathbb{R}$ in Eq. (\ref{poincare_cartan_invariant}).

This construction of an extended phase-space that encompasses the tangent bundle of a four dimensional configuration space allows, not surprisingly, to study relativistic systems using the Hamiltonian formalism. This is particularly useful for us, since the noncommutative relations of NCQM can be encoded as a deformation of the standard symplectic form defined on the phase-space for a single particle. 

\subsection{Canonical transformations and Lorentz transformations}

Consider the diffeomorphisms $\phi: T^*M_1\rightarrow T^*M_1$ that preserve the symplectic form $\omega_e$: these are defined as canonical transformations. The action of the pullback, $\phi^*$, on the symplectic form is therefore $\phi^*\omega_e=\omega_e$. Since any symplectic form is, at least locally, exact (i.e., $\omega_e=\mathrm{d}\theta_e$), one gets that the pullback of $\theta_e$ through $\phi$ must differ from $\theta_e$ itself by at most an exact 1-form, i.e. $\mathrm{d}F_1$. Therefore, this function $F_1$ generates the coordinate transformations that preserve the symplectic structure of the manifold. It is possible to find a generating function, call it $F_2$, whose associated coordinate transformations are precisely the Lorentz transformations of the four-position and four-momentum vectors \cite{Struckmeier_2005}, thus allowing for the construction of Lorentz invariant Hamiltonians in the extended phase-space. Additionally, these transformations are not separable into a spatial coordinate transformation and a time scale change, thus having no analogous transformation in the regular phase-space. One can then take advantage of these transformations to find extensions for some non-relativistic Hamiltonians, which become relativistic in the extended phase-space, and getting the relativistic Hamiltonian on $T^*M$ using $\Psi^{-1}$.

\subsection{Relativistic free particle}

It is particularly useful to understand the discussed formalism, to consider the free particle non-relativistic Hamiltonian, given by:
\begin{equation}\label{nrhamiltonian}
H_{NL}=\frac{1}{2m}p^2,
\end{equation}
which is defined in $T^*M$. One must now find an extension of $H_{NL}$ such that it is Lorentz invariant in $T^*M_1$. This is straightforward and is given by:
\begin{equation}\label{extHamiltonian}
H_1=\frac{1}{2m}\left(p^2-\frac{\mathcal{H}^2}{c^2}\right)+\frac{1}{2}mc^2,
\end{equation}
where $\mathcal{H}$ is the deformation that extends the Hamiltonian to the extended space.
This is properly defined in $T^*M_1$ and since it only depends on $(p^1)^2$ it is Lorentz invariant. The additional $mc^2$ term was added in order to have $H_L=mc^2$ when $p=0$ in the regular phase-space. One should notice that the extended Hamiltonian $H_1$ is constructed as an extension of $H_{NL}$, in the sense that $H_{1|_{\mathcal{H}=0}}=H_{NL}+\mathrm{constant}$. However, it is not a straightforward extension in the sense that, after the reduction process we recover $H_{NL}$: the resulting Hamiltonian should be Lorentz invariant. Indeed, if one now applies $\Psi^{-1}$ to the system $(T^*M_1,\omega_e,H_1)$ it yields $(T^*M,\omega,H_L)$ with $H_L$ being obtained by solving $H_1=0$ and replacing $\mathcal{H}$ by $H_L$, which leads to:
\begin{equation}
\begin{split}
0=&\frac{1}{2m}\left(p^2-\frac{H_L^2}{c^2}\right)+\frac{1}{2}mc^2 \\
\Leftrightarrow H_L^2=&p^2c^2+m^2c^4,
\end{split}
\end{equation}
which is the Hamiltonian for a relativistic free particle and yields the dispersion relation:
\begin{equation}
E^2=p^2c^2+m^2c^4   
\end{equation}

\subsection{Extended Phase-space and Noncommutativity}

In order to find the deformation on the dispersion relation for a single relativistic particle due to the introduction of the additional commutation relations, one uses the extended-phase space formalism, as it is suitable to describe the phase-space of a relativistic particle. However, before the reduction process, one deforms the symplectic form $\omega_e$ in order to accommodate the additional commutation relations, and subsequently one has to find an appropriate coordinate change to bring the deformed form to its canonical form. For this purpose, one can make use of the Darboux's theorem, which guarantees that locally, any symplectic form is reducible to the standard one, by a change of coordinates.

\section{Noncommutative Quantum Mechanics - symplectic structure}

NCQM can be cast in the phase-space formalism of QM by deforming the symplectic form with the extra noncommutative parameters. This implies that the energy, i.e., Hamiltonian, will be different in this context. Before studying the consequences of this deformation, one needs to know how the symplectic form is affected by the introduction of the extra commutation relations. First one considers that the commutators are related with the symplectic form through:
\begin{equation}
\left[f,g\right]=\mathrm{i}\hbar\omega\left(X_f,X_g\right),
\end{equation}
where $f$ and $g$ are functions of position and momentum and $X_{f/g}$ are the vector fields associated with the corresponding function. This vector field is defined through the relation:
\begin{equation}\label{function_vector_field}
\iota_{X_f}\omega=-\mathrm{d}f,
\end{equation}
where $\iota_X$ is the interior product of a vector field and a differential form, $\iota_X:\Omega^p\rightarrow\Omega^{p-1}$, defined by $(\iota_X\omega)(X_1,\dots,X_n)=\omega(X,X_1,\dots,X_n)$. In components:
\begin{equation}
\omega_{ij}X_f^i=-\partial_jf\Leftrightarrow X_f^i=-\omega^{ij}\partial_jf,
\end{equation}
where $\omega^{ij}:=\left(\omega^{-1}\right)_{ij}$ represents the components of the inverse of $\omega$. When applied to the commutator this yields:
\begin{equation}
\left[f,g\right]=\mathrm{i}\hbar\omega_{ij}X_f^iX_g^j=\mathrm{i}\hbar\omega_{ij}\omega^{ik}\omega^{jl}\partial_kf\partial_lg=\mathrm{i}\hbar\omega^{kl}\partial_kf\partial_lg.
\end{equation}
In particular, this implies that:
\begin{equation}
\left[\tilde{z}^i,\tilde{z}^j\right]=\mathrm{i}\hbar\omega^{ij}.
\end{equation}
Therefore, the commutation relations determine uniquely the components of the inverse of the symplectic form. Given this result, one is able to write it in matrix form for the commutative scenario as:
\begin{equation}\label{symplectic_inverse_C}
\omega^{-1}=
\begin{pmatrix}
\boldsymbol{0} & \boldsymbol{\mathrm{Id}} \\
-\boldsymbol{\mathrm{Id}} & \boldsymbol{0}
\end{pmatrix},
\end{equation}
and for the noncommutative one has:
\begin{equation} \label{symplectic_inverse_NC}
\omega_{NC}^{-1}=
\begin{pmatrix}
\frac{\boldsymbol{\Theta}}{\hbar} & \boldsymbol{\mathrm{Id}} \\
-\boldsymbol{\mathrm{Id}} & \frac{\boldsymbol{\mathrm{N}}}{\hbar}
\end{pmatrix},
\end{equation}
where $\boldsymbol{\mathrm{Id}}$ is the $n\times n$ identity matrix and $\boldsymbol{\Theta}$ and $\boldsymbol{\mathrm{N}}$ are skew-symmetric matrices encoding noncommutativity with constant entries $\theta$ and $\eta$ respectively (see e.g. Ref. \cite{Bastos_2008}).
The symplectic form is then given by:
\begin{equation}
\omega_{NC}=\left(1-\frac{\eta\theta}{4\hbar^2}\right)\left[\left(4\hbar^2-\eta\theta\right)\mathrm{d}q^i\wedge\mathrm{d}k_i-4\eta\hbar\mathop{\sum_{i,j=1}}_{i<j}^{3}\mathrm{d}q^i\wedge\mathrm{d}q^j-4\theta\hbar\mathop{\sum_{i,j=1}}_{i<j}^{3}\mathrm{d}k_i\wedge\mathrm{d}k_j\right].
\end{equation}

In this setup, one is able to find a global coordinate transformation that reduces the $\omega_{NC}$ to the standard form. Let the new coordinates be $(x^i,p_i)$. The symplectic form is then written as:
\begin{equation}
\omega=-\mathrm{d}x^i\wedge\mathrm{d}p_i,
\end{equation}
or in the matrix form as $\omega^{-1}$ in Eq.~(\ref{symplectic_inverse_C}).
As seen, in Ref. \cite{Bastos_2008}, a suitable change of coordinates for this task is given by:
\begin{equation}\label{SWmap}
q^i=x^i-\frac{\theta}{2\hbar}\epsilon^{ij}p_j, \qquad k_i=p_i+\frac{\eta}{2\hbar}\epsilon_{ij}x^j.
\end{equation}
This coordinate transformation is noncanonical, since it does not preserve the form of $\omega$. In the following section we will explore the consequences of deforming the spatial part of the symplectic form on the dispersion relation for a relativistic particle, using the extended phase-space formalism.

\subsection{Deformed dispersion relation}

We now consider a free non-relativistic particle with the same Hamiltonian given by Eq. (\ref{nrhamiltonian}) and follow the same procedure that defines the extended Hamiltonian, Eq. (\ref{extHamiltonian}). Then, we consider that the symplectic form in these coordinates, $(p_i,q^i)$, is the deformed NC one. Therefore, in order to obtain the dispersion relation for commutative coordinates, the coordinate change in Eq. (\ref{SWmap}) is used. This yields the extended Hamiltonian:
\begin{equation}
H_1=\frac{1}{2m}\left(\left(p_i+\frac{\eta}{2\hbar}\epsilon_{ij}x^j\right)^2-\frac{\mathcal{H}^2}{c^2}\right)+\frac{1}{2}mc^2.
\end{equation}
This change of coordinates is non-canonical, as the form of the symplectic form is changed. This in turn implies that the Lorentz invariance of the final Hamiltonian is broken. Therefore, the introduction of the noncommutative parameters leads to a violation of Lorentz invariance. As a first consequence, this means that the ensuing results are not independent of the observer performing measurements on the energy and speed of particles. The consequences of this fact will be further explored in the next section. 
After the reduction process, the Hamiltonian for a free particle in the deformed phase-space is given by:
\begin{equation}
H_{NC}^2=\left(p_i+\frac{\eta}{2\hbar}\epsilon_{ij}x^j\right)^2c^2+m^2c^4,
\end{equation}
which, interpreting $H_{NC}$ as the measured energy of a free particle, leads to a modified relativistic dispersion relation:
\begin{equation} \label{deformed_dispersion}
E^2=\left(p_i+\frac{\eta}{2\hbar}\epsilon_{ij}x^j\right)^2c^2+m^2c^4.
\end{equation}
In turn, this deformation yields a modification to the propagation of free massless particles, given by:
\begin{equation}\label{c_prime}
\boldsymbol{c'}=\nabla_{\boldsymbol{p}}E=\frac{1}{2E}\nabla_{\boldsymbol{p}}E^2=\frac{c^2}{E}\left(p_i+\frac{\eta}{2\hbar}\epsilon_{ij}x^j\right)\boldsymbol{\hat{x}^j},
\end{equation}
which leads 
to the measurable speed of light as $\|\boldsymbol{c'}\|=c$. Therefore, even with the deformation in the dispersion relation, the speed of light is maintained.

\section{Testing the deformation of the dispersion relation}

In order to assess the validity of the deformation introduced in the dispersion relation, and since Lorentz invariance is no longer a symmetry of the system, one should be able to test quantitatively our result using tests of Lorentz invariance. Given that the resulting modification is dependent on the distance from the photon to the observer, gamma ray burst (GRB) tests of Lorentz invariance are suitable to examine this deformation as the distance scale is sufficiently large for the effects to be detectable. Furthermore, as the observations of these phenomena are made from Earth, it is convenient to use spherical coordinates for the deformed dispersion relation. In these coordinates, the dispersion relation becomes:
\begin{equation}
E^2=c^2p^2+\frac{c^2\eta}{\hbar}r\left[\frac{\eta}{2\hbar}r f\left(\theta,\phi\right)+p_\phi\left(\sin\phi+\cos\phi\right)+p_\theta g\left(\theta,\phi\right)\right],
\end{equation}
where $f\left(\theta,\phi\right)=(1/2)(2-\cos\phi\sin2\theta+\sin 2 \theta\sin\phi+\sin^2\theta\sin 2\phi)$ and $g\left(\theta,\phi\right)=\cos\theta\cos\phi-\cos\theta\sin\phi+\sin\phi-\sin\theta$. One immediately notices that the energy of the particle is not independent from $\theta$ and $\phi$, which clearly breaks isotropy. This is not surprising since the commutation relations also break this property. If one assumes that the observed photons carry only radial momentum their energy is then given by:
\begin{equation}
E^2=c^2p^2+\frac{c^2\eta^2}{2\hbar^2}r^2 f\left(\theta,\phi\right).
\end{equation}
This is a reasonable assumption, as the distance between the source and the receiver is very large compared to the radius of the Earth. In order to test the obtained expression for the energy of massless particles, an additional assumption is made, namely that the angular factor is a minor contribution to the energy deformation. One could also argue that phenomenologically the distribution of GRBs in the sky is isotropic. With this assumption, the dispersion relation becomes:
\begin{equation}
\begin{split}
E&\simeq cp\sqrt{1+\frac{\eta^2}{2\hbar^2}\frac{r^2}{p^2}}\simeq cp\left(1+\frac{\eta^2}{4\hbar^2}\frac{r^2}{p^2}\right) \\
\Leftrightarrow \frac{\Delta E}{E}&\simeq\frac{\eta^2}{4\hbar^2}\frac{r^2}{p^2}.
\end{split}
\end{equation}
One is now able to use GRB data to estimate the magnitude of the noncommutative parameter $\eta$.
In order to do so, one assumes that the noncommutative correction is at most given by the uncertainty value in the measurement of the GRB energy. Therefore one needs the GRB energy, its uncertainty and the distance from Earth. Additionally, one uses that $E=pc+O(\eta)$, and $O(\eta)\ll 1$, to get:
\begin{equation} \label{eta_determination}
\frac{\Delta E}{E}=\frac{\eta^2c^2}{4\hbar^2}\frac{r^2}{E^2} \Leftrightarrow \eta=\frac{2\hbar}{cr}\sqrt{E\Delta E}.
\end{equation}

Hence, one needs to know the detected energy of the emitted photons, its uncertainty, as well as the distance of the GRB. Data with this level of detail is not available at the moment, as far as the authors are aware. Instead, one uses average values for the energy in a certain detected band as a substitute for the photon values of these quantities. In order to calculate the distance of the GRB, the redshifts of these objects are used and the light-travel distance is computed. Data of some selected GRBs \cite{database,database_2} is presented in Table \ref{GRBdata}.

\begin{table}
\caption[]{Gamma Ray Burst data \cite{database,database_2} and upper bound on $\sqrt{\eta}$}
\label{GRBdata}
\centering
\begin{tabular}{m{2cm} c c m{2cm} m{2cm} c}
\hline
 & Fluence ($F$) (erg/cm$^2$) & $\Delta F {\small (\times 10^{-8}}$  erg/cm$^2$) & $\qquad z$ & $r\,(\times 10^{25}\mathrm{m})$ & $\sqrt{\eta}$ (eV) \\[4pt]
\hline
180703A & $7.8096\times 10^{-6}$ & $2.3435$ & $\,\,$ 0.6678 & 5.86083 & $4.252\times 10^{-12}$ \\
171010A & $3.3315\times 10^{-4}$ & $5.1486$ & $\,\,$ 0.3285 & 3.50175 & $1.712\times 10^{-11}$ \\
170214A & $8.2807\times 10^{-5}$ & $6.3294$ & $\,\,$ 2.53 & 10.5876 & $7.319\times 10^{-12}$ \\
160625B & $2.7614\times 10^{-4}$ & $9.7813$ & $\,\,$ 1.406 & 8.71085& $1.216\times 10^{-11}$ \\
160623A & $2.1145\times 10^{-6}$ & $3.9158$ & $\,\,$ 0.367 & 3.82187 & $4.319\times 10^{-12}$ \\
160509A & $8.7354\times 10^{-5}$ & $8.0344$ & $\,\,$ 1.17 & 8.02423 & $9.044\times 10^{-12}$ \\
150514A & $2.5347\times 10^{-6}$ & $2.4576$ & $\,\,$ 0.807 & 6.58455 & $3.064\times 10^{-12}$ \\
141028A & $1.7317\times 10^{-5}$ & $5.1850$ & $\,\,$ 2.33 & 10.3613 & $4.760\times 10^{-12}$ \\
140808A & $2.1951\times 10^{-6}$ & $1.6704$ & $\,\,$ 3.29 & 11.2211 & $2.056\times 10^{-12}$ \\
140801A & $8.7936\times 10^{-6}$ & $2.1775$ & $\,\,$ 1.32 & 8.47863 & $3.576\times 10^{-12}$ \\
140623A & $1.6891\times 10^{-6}$ & $2.5486$ & $\,\,$ 1.92 & 9.7792 & $2.293\times 10^{-12}$ \\
140620A & $3.4297\times 10^{-6}$ & $3.3173$ & $\,\,$ 2.04 & 9.96892 & $2.895\times 10^{-12}$ \\
140606B & $3.7519\times 10^{-6}$ & $2.4159$ & $\,\,$ 0.384 & 3.95833 & $4.341\times 10^{-12}$ \\
140508A & $3.0887\times 10^{-5}$ & $6.2225$ & $\,\,$ 1.027 & 7.5222 & $6.757\times 10^{-12}$
\end{tabular}
\end{table}

Distances are computed from the redshift of associated counterparts in other wavelengths using the standard $\Lambda$-CDM model, with parameters $H_0=70 \,km \,s^{-1} Mpc^{-1}$, $\Omega_m=0.27$, $\Omega_\Lambda=0.73$ and $k=0$, and is given by:
\begin{equation}
d(z)=d_H\int_0^z\frac{\mathrm{d}z'}{(1+z')E(z')},
\end{equation}
where $d_H=c/H_0$ is the Hubble distance and 
\begin{equation}
E(z)=\sqrt {\Omega _r(1+z)^{4}+\Omega _m(1+z)^{3}+\Omega _k(1+z)^{2}+\Omega _\Lambda},
\end{equation}
with $H(z)=H_0E(z)$ being the Hubble factor at redshift $z$. 
The energy is taken to be related to the fluence measured in the BATSE standard $50-300$ keV energy band. Since the fluence is given by $F=E/A$, where $A$ is the cross sectional area of detection, Eq.~(\ref{eta_determination}) changes to:
\begin{equation}
\eta=\frac{2\hbar A }{c r}\sqrt{F\Delta F}.
\end{equation}
With the data above \cite{database,database_2}, the estimate on $\eta$ is given, up to an area order of magnitude. This estimations are shown in Table \ref{GRBdata} for $A=1m^2$.

These results allows us to conclude that, if we take the uncertainty in the GRB energy as due to the deformed dispersion relation, given in Eq.~(\ref{deformed_dispersion}), the magnitude of the fundamental momentum scale introduced by NCQM, $\sqrt{\eta}$, would be of the order of $\sim 10^{-12}$ eV/c for $A\sim 1\, \mathrm{m}^2$. This is a very stringent upper bound, improving on previous results where low-energy tests of Lorentz invariance were used, namely $\sqrt{\eta}\lesssim 10^{-5} \mathrm{eV/c}$ \cite{Queiroz_2011, Leal_2015}.

\section{Riemannian metric on noncommutative phase-space}

It is useful for the understanding of the deformed dispersion relation, Eq.~(\ref{deformed_dispersion}), to introduce a Riemannian metric in the phase-space, or more precisely, in the extended phase-space. In order to do this, one first considers a smooth manifold equipped with a differential 2-form, $\omega$. This is a symplectic manifold if $\omega$ is closed and nondegenerate. This 2-form $\omega$ is called a symplectic 2-form. This structure is the basis of the phase-space.

The usual symplectic form defined on phase-space is $\omega=-\mathrm{d}x^i\wedge\mathrm{d}p_i$, $i=1,\dots,n$, which in matrix form is given by Eq.~(\ref{symplectic_original}).
In addition to the symplectic form, one can introduce an almost complex structure (ACS) on $M$. This is a smooth (1,1)-rank tensor, $J$ such that $J^2=-\mathrm{Id}$. A manifold equipped with an ACS is called an almost complex manifold.

Given a symplectic manifold (M, $\omega$), it is always possible to find an ACS in such a way that a metric is well defined on the manifold \cite{Blair_book}. In order to do so, one must find a compatible ACS. This condition is satisfied if:
\begin{equation} \label{ACS}
\omega\left(J\left(X_{z^i}\right),J\left(X_{z^j}\right)\right)=\omega\left(X_{z^i},X_{z^j}\right),
\end{equation}
where $X_{z^i}$ is the vector field associated with the function $z^1$, defined by Eq. (\ref{function_vector_field}). This condition thus requires that the ACS preserves the action of the symplectic form on vector fields in $TM$. Let us now consider the structure (M, $\omega$, J) as a symplectic manifold equipped with an ACS. Then, it is always possible to define a Riemannian metric on M, i.e., a symmetric rank (0,2)-tensor, given by:
\begin{equation}
g\left(X_{z^i},X_{z^j}\right):=\omega\left(X_{z^i},J(X_{z^j})\right).
\end{equation}
These objects form a compatible triplet, in the sense that defining two of these structures allows for specifying the third one in a unique way.

\subsection{Commutative phase-space}

For the commutative phase-space, the symplectic form is given by:
\begin{equation}
\boldsymbol{\omega}=-\mathrm{d}x^i\wedge\mathrm{d}p_i,
\end{equation}
$i=1,...,n$. The ACS satisfying Eq. (\ref{ACS}) acts on the coordinate vector fiels as:
\begin{equation}
\boldsymbol{J}(X_{x^i})=X_{p_i}, \quad \boldsymbol{J}(X_{p_i})=-X_{x^i}
\end{equation}
The induced compatible Riemannian metric is then $\boldsymbol{g}=\boldsymbol{\mathrm{Id}}_{2n}$.

\subsection{Noncommutative phase-space}

The deformed symplectic for of the NC phase-space is easily characterized by its inverse as seen in Eq~(\ref{symplectic_inverse_NC}). The ACS action on the vector fields $X_{\tilde{z}^i}$ is given by:
\begin{equation} \label{NC_ACS}
\begin{split}
\boldsymbol{J}(X_{q^i})=\sqrt{\frac{\theta}{\eta}}X_{k_i}\\
\boldsymbol{J}(X_{k^i})=-\sqrt{\frac{\eta}{\theta}}X_{x^i}
\end{split}
\end{equation}
In this scenario, the metric of the compatible trio acts on the above vector fields as:
\begin{equation} \label{psmetric}
\begin{split}
\boldsymbol{g}\left(X_{q^i},X_{q^j}\right)=\sqrt{\frac{\theta}{\eta}}\delta_{ij} \\
\boldsymbol{g}\left(X_{k_i},X_{k_j}\right)=\sqrt{\frac{\eta}{\theta}}\delta_{ij} \\
\boldsymbol{g}\left(X_{q^i},X_{k_j}\right)=-\sqrt{\frac{\eta\theta}{\hbar^2}}\epsilon_{ij} \\
\boldsymbol{g}\left(X_{k_i},X_{q^j}\right)=\sqrt{\frac{\eta\theta}{\hbar^2}}\epsilon_{ij}.
\end{split}
\end{equation}
The metric components are related to its value on the vector fields $X_{\tilde{z}^i}$ through:
\begin{equation}
g_{ij}=\omega_{ik}\omega_{jl}g\left(X_{\tilde{z}^k},X_{\tilde{z}^l}\right),
\end{equation}
which, for $n=2$, i.e., a 4 dimensional phase-space, yields in matrix form:
\begin{equation}
\boldsymbol{g}=
\begin{pmatrix}
\hbar^2\frac{\sqrt{\frac{\eta}{\theta}}\hbar^2-\sqrt{\eta^3\theta}}{(\theta\eta-\hbar^2)^2} & 0 & 0 & \frac{\hbar\sqrt{\eta\theta}}{\hbar^2-\theta\eta} \\
0 & \hbar^2\frac{\sqrt{\frac{\eta}{\theta}}\hbar^2-\sqrt{\eta^3\theta}}{(\theta\eta-\hbar^2)^2} & -\frac{\hbar\sqrt{\eta\theta}}{\hbar^2-\theta\eta} & 0 \\
0 & -\frac{\hbar\sqrt{\eta\theta}}{\hbar^2-\theta\eta} & \hbar^2\frac{\sqrt{\frac{\theta}{\eta}}\hbar^2-\sqrt{\eta\theta^3}}{(\theta\eta-\hbar^2)^2} & 0 \\
\frac{\hbar\sqrt{\eta\theta}}{\hbar^2-\theta\eta} & 0 & 0 & \hbar^2\frac{\sqrt{\frac{\theta}{\eta}}\hbar^2-\sqrt{\eta\theta^3}}{(\theta\eta-\hbar^2)^2}
\end{pmatrix}.
\end{equation}
The volume form associated with this metric is:
\begin{equation}\label{volume_form_metric_NC}
\mathrm{volume}_g=\sqrt{|g|}\mathrm{d}^4\tilde{z}=\frac{\hbar^4}{\hbar^2-\eta\theta}\mathrm{d}^4\tilde{z}.
\end{equation}
This coincides with the induced volume by the symplectic form:
\begin{equation}
\mathrm{volume}_\omega=\omega\wedge\omega=\frac{\hbar^4}{\hbar^2-\eta\theta}\mathrm{d}^4\tilde{z},
\end{equation}
which is to be expected as the manifold is K$\ddot{\mathrm{a}}$hler. One should notice that it is not possible to take the naive commutative limit, i.e. $\theta,\eta\rightarrow 0$, since the metric becomes ill-defined. A way around is to consider $\theta/\eta=\alpha$ and then take the above limit with $\alpha$ fixed to get:
\begin{equation}
\boldsymbol{g}=\mathrm{diag}\left(\sqrt{\alpha^{-1}},\sqrt{\alpha^{-1}},\sqrt{\alpha},\sqrt{\alpha}\right),
\end{equation}
which yields the commutative limit iff $\alpha=1$, that is $\theta=\eta$.
\subsection{Extended Noncommutative phase-space}

Let us apply the former procedure to the extended symplectic form, Eq.~(\ref{extended_sf}).
The action of the ACS on the vector fields of the spatial coordinates is the same as the one given in Eq.~(\ref{NC_ACS}), while its action on the time and energy vector fields is the noncommutative one, i.e.,
\begin{equation}
J(X_t)=X_\mathcal{H}, \quad J(X_\mathcal{H})=-X_t.
\end{equation}
We are then able to define the extended noncommutative phase-space metric by its action on the coordinate vector fields as before. Its action on the spatial vectors is the same as in Eq.~(\ref{psmetric}) and on the remaining vectors it is given by:
\begin{equation}
g(X_t,X_t)=g(X_\mathcal{H},X_\mathcal{H})=-1.
\end{equation}
From this we get that the metric components are given by:
\begin{equation}
\boldsymbol{g}=
\left(
\begin{matrix}
 -1 & 0 & 0 & 0 & 0 & 0 \\
 0 & \hbar^2\frac{\sqrt{\frac{\eta}{\theta}}\hbar^2-\sqrt{\eta^3\theta}}{(\theta\eta-\hbar^2)^2} & 0 & 0 & 0 & \frac{\hbar\sqrt{\eta\theta}}{\hbar^2-\theta\eta} \\
 0 & 0 & \hbar^2\frac{\sqrt{\frac{\eta}{\theta}}\hbar^2-\sqrt{\eta^3\theta}}{(\theta\eta-\hbar^2)^2} & 0 & -\frac{\hbar\sqrt{\eta\theta}}{\hbar^2-\theta\eta} & 0 \\
 0 & 0 & 0 & -1 & 0 & 0 \\
 0 & 0 & -\frac{\hbar\sqrt{\eta\theta}}{\hbar^2-\theta\eta} & 0 & \hbar^2\frac{\sqrt{\frac{\theta}{\eta}}\hbar^2-\sqrt{\eta\theta^3}}{(\theta\eta-\hbar^2)^2} & 0 \\
 0 & \frac{\hbar\sqrt{\eta\theta}}{\hbar^2-\theta\eta} & 0 & 0 & 0 & \hbar^2\frac{\sqrt{\frac{\theta}{\eta}}\hbar^2-\sqrt{\eta\theta^3}}{(\theta\eta-\hbar^2)^2} \\
\end{matrix}
\right),
\end{equation}
with volume form being the same as in Eq.~(\ref{volume_form_metric_NC}), which again coincides with the volume form induced by $\omega_e$. Likewise in the non-extended NC phase-space, the limit $\eta,\theta\rightarrow 0$ is not well defined unless $\theta=\eta$.

The implications of this metric structure will be discussed elsewhere.

\section{Conclusions}

In this work the deformation of the relativistic dispersion relation arising from the breaking of Lorentz symmetry in phase-space NCQM is analysed. For this purpose, the framework of phase-space as a symplectic manifold is used, as well as its extension to include time as a coordinate and not as a parameter. This leads to the extended phase-space formalism. The deformation of the HW algebra of observables is used to compute the deformation in the symplectic form defined in the extended phase-space. This gives rise to fundamental length and momentum scales, that results in the breaking of Lorentz invariance. This fact, together with the symplectic point of view for the phase-space allows for the use of Darboux's map to find the deformation of the dispersion relation for a relativistic particle. The new relation is tested with GRB data from several sources at different redshifts and the obtained constraint for the momentum scale is $\sqrt{\eta}\lesssim 10^{-12}\, \mathrm{eV/c}$. This is a very stringent bound and improves on previous low-energy physics results, $\sqrt{\eta}\lesssim 10^{-5}\, \mathrm{eV/c}$ \cite{Queiroz_2011,Leal_2015}.

Additionally, a pseudo-Riemannian metric is introduced in the extended noncommutative phase-space, via a compatible triplet comprised of a symplectic form, an almost complex structure and a metric tensor. This allows for a straightforward notion of distance in phase-space that arises naturally from compatibility requirements on these three structures. With this framework, the noncommutative metric tensor is constructed. It is found that the naive commutative limit $\theta,\eta\rightarrow 0$ cannot be taken since it renders the metric ill defined. Rather, this limit must be considered for fixed $\theta/\eta$ ratio. With this, one recovers the commutative limit iff $\theta=\eta$. The significance of this result, as well as its implications will be discussed elsewhere.

\vspace{.5 cm}
{\em Acknowledgments} -- The work of PL is supported by FCT (Portuguese Fundaçâo para Ciência e Tecnologia) grant PD/BD/135005/2017. The work of OB is partially supported by the COST action MP1405.

\end{document}